\documentclass[12pt]{article}

\usepackage{epsfig,a4wide}

\newcommand{\Journal}[2]{#1, #2}
\newcommand{\Author}[2]{#1. #2}
\newcommand{\Title}[5]{{\it #1} #2 {\bf #3}, #4, (#5)}

\newcommand{\TMF}{Teor. Mat. Fiz.}
\newcommand{\TMP}{Theor. Math. Phys.}
\newcommand{\ZETF}{Zh. Eksp. Teor. Phys.}
\newcommand{\AMEP}{Appl. Math. \& Eng. Phys.}
\newcommand{\EWES}{Electomagn. W. \& Electr. Sys.}

\newcommand{\QM}{Quantum Mechanics - Non-relativistic Theory}
\newcommand{\PNF}{Physical Nature of Fireball}

\newcommand{\PK}{Physical Kinetics}

\title{Electron gas oscillations in plasma.\\Theory and applications.}

\author{Maxim~Dvornikov\thanks{Department of Theoretical Physics,
Moscow State University, 119992 Moscow, Russia; E-mail address:
maxim\_dvornikov@aport.ru} \and
Sergey~Dvornikov\thanks{N.~N.~Andreev Institute of Acoustics,
'Shvernika' Str 4, 117036 Moscow, Russia; E-mail address:
sergedvo@mail.ru}}

\begin{document}

\maketitle

\begin{abstract}
We analyze the obtained solutions of the non-linear Shr{\"o}dinger
equation for spherically and axially symmetrical electrons density
oscillations in plasma. The conditions of the oscillations process
existence are examined. It is shown that in the center or on the
axis of symmetry of the systems the static density of electrons
enhances. This process results in the increasing of density and
pressure of the ion gas. We suggest that this mechanism could
occur in nature as rare phenomenon called the fireball and could
be used in carrying out the research concerning controlled fusion.
The description of the experiments, carried out for the purpose to
generate long-lived spherical plasma structures, is presented.
\end{abstract}

\section{\label{INT} Introduction}

\begin{figure}[htb]
\begin{center}
\epsfig{file=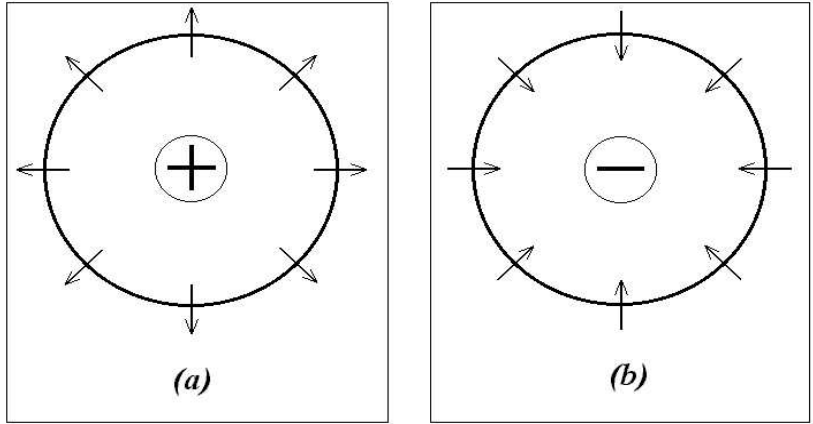,height=6cm,width=11cm} \caption{\label{gosc}
the process of electron gas oscillations in plasma, with electrons
moving from (a) and to (b) the central area.}
\end{center}
\end{figure}

The studying of electron gas oscillations in plasma is interesting
physical problem, mainly because it enables one to examine plasma
properties from the theoretical point of view as well as it
provides the basis for subsequent experiments in this area. Let us
discuss the electric charge variation in electroneutral plasma. If
a volume charge appears in such a system, i.e. electrons density
increases or decreases in some finite area, then, after an
external influence is over, the oscillating process consisting in
periodical changes of the sign of the considered volume charge is
known to appear. We can roughly neglect the motion of positively
charged ions since their mass is several orders of magnitude
greater then electron mass. The process of electron gas
oscillations is schematically shown in Fig.~\ref{gosc}. The
electrons motion from the central area is presented in
Fig.~\ref{gosc}(a). Thus, the central region acquires excessive
positive charge. Electrons moving to the central area are depicted
in Fig.~\ref{gosc}(b). In this case the central region becomes
negatively charged.
It is necessary to remind that the frequency of this process
$\omega_{p}$, called plasma frequency, is related to free
electrons density in plasma $n_{0}$ by the formula
\begin{equation}
  \label{plfr}
  \omega _{p}^{2} =
  \frac{{4 \pi e^{2}n_{0}}}{{m}},
\end{equation}
where $e$ and $m$ are the charge and the mass of the electron.

Numerous attempts were made to describe electron gas oscillations
within the context of both classical and quantum mechanics (see,
for example, Ref.~\cite{Li.Pi/PK(79)}). The method of kinetic
equation is one of the most commonly used approaches to this
problem in frames of classical theory. The kinetic equation for
each type of particles has the form:
\begin{equation}
  \label{kineq}
  \frac{\partial f_{\alpha}}{\partial t}+
  {\bf v}\frac{\partial f_{\alpha}}{\partial {\bf r}}+
  {\dot {\bf p}}\frac{\partial f_{\alpha}}{\partial {\bf p}}=
  {\rm St}f_{\alpha},
\end{equation}
where $f_{\alpha}({\bf r},{\bf p},t)$ is the distribution function
depending on the coordinate ${\bf r}$, momentum ${\bf p}$ and time
$t$ for either electron $(\alpha=e)$ or ion $(\alpha=i)$
component, ${\rm St}f_{\alpha}$ is the collision integral. The
derivative ${\dot {\bf p}}$ is equal to $q\left({\bf
E}+\frac{1}{c}[{\bf v}\times{\bf B}]\right)$ for the case of
electromagnetic interactions, $q$ is the particle charge.

The collision integral is negligible if the mean free path $l$ of
a particle is much greater than the distance of the
electromagnetic field (produced by charged particles in plasma)
variation $L$: $l\gg L$. The Eq.~(\ref{kineq}) with ${\rm
St}f_{\alpha}=0$ for $\alpha=i,e$ should be supplied with the
system of averaged Maxwell equations in order for the electric
${\bf E}$ and magnetic ${\bf B}$ fields to be specified. The
electromagnetic field, determined in such a manner, is called the
self-consistent field. The self-consistent field conception
enables one to investigate electromagnetic properties of plasma
and describe various plasma phenomena \cite{Li.Pi/PK(79)}. The
kinetic equation with ${\rm St}f_{\alpha}\not=0$ takes into
account the collision processes between particles in plasma. There
exist several types of collision integrals \cite{Li.Pi/PK(79)}.

There is, however, another approach, developed in
Refs.~\cite{Dr.Ku/TMP(96),Ma.Ku/TMP(99)} to describe the evolution
of $N$ interacting particles. It is known that the classical
dynamics of $N$ interacting particles can be presented by the
system of differential equations of motion in configurational
space of $3N$ dimensions or in $6N$-dimensional phase space. It is
also possible to describe the evolution of such a system by means
of partial differential equations in three-dimensional physical
space as the dynamics of microscopic singular material fields
(see~Ref.~\cite{Dr.Ku/TMP(96)}). A particle in this approach was
determined as $\delta$ function distribution containing spatial
coordinates and time independently. This method allows one to
implement the mapping of trajectories dynamics in the fields
dynamics and vice versa. To describe the particles dynamics on the
macroscopic scale in this approach one should smooth the
microscopic distributions. This procedure is well known in
classical electrodynamics and hydrodynamics.

The transfer to the quantum-mechanical description is realized by
replacing the dynamic functions with Hermitian operators. In this
case the dimensionality of the configuration space is conserved
and the state of the system is completely defined by the wave
functions in $3N$-dimensional space. However, physical
characteristics of the $N$ particles system are determined in
three-dimensional physical space.

The quantum-mechanical description of the system of $N$ charged
particles with arbitrary masses was considered in
Ref.~\cite{Ma.Ku/TMP(99)}. Particles were taken to interact by
Coulomb forces and with external classical electromagnetic field
characterized by vector ${\bf A}({\bf r},t)$ and scalar
$\varphi({\bf r},t)$ potentials. It is worth mentioning that the
evolution of $N$ particles was described in three-dimensional
physical space. The exact microscopic equations for the
description of the quantum dynamics of the $N$ particles system in
physical space were obtained in that paper. The transfer to the
macroscopic observable fields was implemented by means of
smoothing procedure of microscopic functions. Note that the
quantum description of the $N$ particles system can be formally
reduced to one particle Shr{\"o}dinger equation if we express all
operators in terms of creation and annihilation operators.
However, one particle wave functions are the operators in this
approach. In Ref.~\cite{Ma.Ku/TMP(99)} it was shown that the
quantum dynamics of the $N$ particles system can be described
using non-operator wave function in three-dimensional physical
space. This result is of great importance.

We applied the formalism developed in Ref.~\cite{Ma.Ku/TMP(99)} to
the description of the electron gas oscillations in plasma. The
solutions of linearized Shr{\"o}dinger equation in the
approximation of the self-consistent field were obtained in
Ref.~\cite{Dv.Dv.Sm/AMEP(01)}.

The main goal of this paper is to study electron gas oscillations
in plasma using quantum mechanical approach. The quantum
description is implemented in three-dimensional physical space. In
Sec.~\ref{TH} we analyze spherically and axially symmetrical
solutions of the non-linear Shr{\"o}dinger equation. The
conditions of the oscillations process existence are examined. It
is found that in the center or on the axis of symmetry of the
systems the static density of electrons enhances. This process
leads to the increasing of density and pressure of the ion gas.
Then, in Sec.~\ref{APP} we discuss possible applications of the
considered model. We suggest that this mechanism could occurs in
nature as rare phenomenon called the {\it fireball} and could be used
in carrying out the research concerning controlled fusion. The
description of the experiments, carried out for the purpose to
generate long-lived spherical plasma structures, is presented.
Finally, in Sec.~\ref{CONCL} we discuss our results.

\section{\label{TH} Quantum description of electron gas oscillations}

We will assume electrons in plasma to be a quantum many body
system. Such an assumption is due to the fact that, as it will be
shown below, the obtained solutions have characteristic sizes of
atomic order. The complex $\Psi$ function is introduced in
three-dimensional space and has the form
\begin{equation}
  \label{wfun}
  \Psi \left( {{\bf r},t} \right) =
  \sqrt {n_{e} \left( {{\bf r},t} \right)}
  e^{\frac{{i}}{{\hbar} }\sigma  \left({\bf r},t \right)},
\end{equation}
where $n_{e}({\bf r},t)=\left| \Psi \right|^{2}$ is the density of
electrons, $\sigma ({\bf r},t)$ is the phase of the function
$\Psi$. The function $\Psi$ satisfies the partial differential
equation:
\begin{equation}
  \label{difeq}
  i\hbar \frac{{\partial \Psi
  }}{{\partial  t}} =
  {\hat H} \Psi .
\end{equation}
The Hamiltonian in the Eq.~(\ref{difeq}) is expressed in the
following way \cite{Ma.Ku/TMP(99),Dv.Dv.Sm/AMEP(01)}
\begin{equation}
  \label{ham}
  {\hat H} =
  \frac{1}{2m}
  \left(
  \frac{\hbar}{i}{\bf \nabla}-\frac{e}{c}{\bf A}({\bf r},t)
  \right)^{2}+
  e\varphi({\bf r},t)+
  e^{2}\int d^{3}{\bf r}^{\prime}
  G({\bf r}-{\bf r}^{\prime})|\Psi|^{2}+
  \theta({\bf r},t),
\end{equation}
where $G\left( {\bf r}-{\bf r}^{\prime} \right)= {\frac{1}{\left|
{\bf r}-{\bf r}^{\prime} \right|}}$. In the Eq.~(\ref{ham}) the
first two terms are the components of single electron Hamiltonian
in the external electromagnetic field. The third term presents the
potential energy of the electron in the self-consistent
electrostatic field created by the whole electrons system, with
number density of particles being equal to $\left| \Psi
\right|^{2}$. The function $\theta \left( {\bf r},t \right)$
describing exchange interactions between electrons has the form
\cite{Ma.Ku/TMP(99),Dv.Dv.Sm/AMEP(01)}
\begin{equation}
  \label{thet}
  \theta  \left( {\bf r},t \right)=
  \frac{{\hbar ^{2}}}{{2m}}\frac{{\Delta \Psi} }{{\left| {\Psi}
  \right|}} + \int\limits_{ {\bf r}_{0}}^{\bf r}
  {\frac{{dp_{e} \left(
  {\bf r},t \right)}}{{\left| {\Psi}  \right|^{2}}}} + e^{2}\int
  {d^{3}{\bf r}^{\prime}\int\limits_{{\bf r}_{0}}^{\bf r}
  {d}G\left( {{\bf r} - {\bf r}^{\prime}} \right)
  \frac{{q_{2} \left( {{\bf r},{\bf r}^{\prime},t} \right)}}{{\left|
  {\Psi}  \right|^{2}}}} ,
\end{equation}
where $p_{e} \left( {\bf r},t \right)$ is the pressure of the
electron gas, $q_{2} \left( {\bf r},{\bf r}^{\prime},t \right)$ is
the correlation function.

Therefore, in order to resolve exactly the considered problem one
should take into account all terms in the Eq.~(\ref{thet}). In
Ref.~\cite{Dv.Dv.Sm/AMEP(01)} we assumed that the contribution of
exchange interactions to the dynamics of free electrons in plasma
is much smaller than the contribution of self-consistent
electrostatic field. This assumption is valid at low density of
free electrons in plasma. Then, in describing the dynamics of the
electron gas we neglected the function $\theta \left( {\bf r},t
\right)$. As it will be seen from further speculations, these
rough approximations allow us to get some characteristics of the
oscillations process which are close enough to those obtained from
the treatment of similar problem within the classical approach.

Let us consider the electroneutral plasma formed by the singly
ionized gas, with the energy of electrons being more than the
ionization potential of this gas. We will suppose that plasma
possesses a spherical symmetry for density and velocities
distribution of the electron and the ion gases. Note that there is
no electromagnetic radiation in this system since the magnetic
field is absent throughout the volume. The magnetic field, and
hence the radiation, can appear if plasma parameters deviate from
the spherically symmetrical distribution. If such deviations are
small, the system is likely to reconstruct its internal structure
and becomes spherically symmetrical again. The problem of the
system stability to such deviations is under investigation now.

Taking into account the small mobility of heavy ions in gas
compared to the mobility of electrons we will suppose that the
density of ions is the constant value $n_{i} \left( {\bf r},t
\right) = n_{0}$. We will also consider that in our case there are
no external electromagnetic fields except those of positively
charged ions. The potential of self-consistent field created by
the electron gas is represented by the formula (using spherical
coordinates):
\begin{equation}
  \label{potU}
  U_{e}=e\int d^{3} {\bf r}^{\prime}
  G({\bf r}-{\bf r}^{\prime})|\Psi|^{2} =
  4{\pi}e\int\limits_{r}^{\infty} \frac{dR}{R^{2}}
  \int\limits_{0}^{R} x^{2} |{\Psi}(x,t)|^{2}dx.
\end{equation}
Similarly, for the potential $\varphi$ of singly ionized gas with
density of ions $n_{i}(r,t)$ one has
\begin{equation}
  \label{potphi}
  \varphi =-4{\pi}e\int\limits_{r}^{\infty} {\frac{dR}{R^{2}}}
  \int\limits_{0}^{R} x^{2} n_{i}(x,t) dx.
\end{equation}
Thus, taking into account
Eqs.~(\ref{potU}) and (\ref{potphi}),
the Eq.~(\ref{difeq}) can be represented in the
following way
\begin{equation}
  \label{trdifeq}
  i\hbar \frac{{\partial \Psi} }{{\partial
  t}} + \frac{{\hbar ^{2}}}{{2m}}\Delta
  \Psi - 4\pi  e^{2}
  \Psi F({{\left|{\Psi} \right|^{2}}}-n_{0})=0,
\end{equation}
where $\Delta={\frac{\partial^{2}}{\partial r^{2}}}+
{\frac{2}{r}}{\frac{\partial}{\partial r}}$ is the Laplas operator
in the spherical coordinate system and
$$
  F(\ldots)=\int\limits_{r}^{\infty} {\frac{dR}{R^{2}}}
  \int\limits_{0}^{R} x^{2} (\ldots) dx.
$$

Moreover, we demand that the system should be electroneutral as a
whole, i.e. the condition must be satisfied:
\begin{equation}
  \label{eln}
  \lim_{R\rightarrow\infty} {\frac{1}{R^{3}}}
  \int\limits_{0}^{R} x^{2} {{\left|{\Psi}(x,t) \right|^{2}}}dx=
  {\frac{n_{0}}{3}}.
\end{equation}
We will search for a solution of the Eq.~(\ref{trdifeq}) in the
form:
\begin{equation}
  \label{form}
  \Psi = \Psi _{0} + \chi (r) e^{ - i\omega t} \quad .
\end{equation}
Here $\Psi _{0}$ is the solution in the case of unperturbed
electroneutral plasma, $|\Psi _{0}|^{2}=n_{0}$, $\chi (r)$ is the
complex function describing small perturbations on frequency
$\omega$, $|\chi|\ll|\Psi_{0}|$. Using the Eq.~(\ref{form}), we
get that $|\Psi|^{2}=n_{0}+\Psi_{0}f+|\chi|^{2}$, where $f=\chi
e^{-i\omega t}+\chi^{*} e^{i\omega t}$. Taking into account that
$|\chi|\ll|\Psi_{0}|$, we obtain that $|\Psi|^{2}\approx
n_{0}+\Psi_{0}f$.

Then, we substitute the function $\Psi$, given in the
Eq.~(\ref{form}), and approximate expression for $|\Psi|^{2}$ in
the Eq.~(\ref{trdifeq}). Having considered the complex-conjugate
equation together with the obtained one, it was easy to get the
following equation for the function $f$:
\begin{equation}
  \label{eqf}
  \hbar \omega  f + \frac{{\hbar ^{2}}}{{2m}}\Delta f - 4\pi
  e^{2}\Psi _{0} \left( {2\Psi _{0} + f} \right) F\left( {f}
  \right)=0.
\end{equation}

\begin{figure}[htb]
\begin{center}
  \epsfig{file=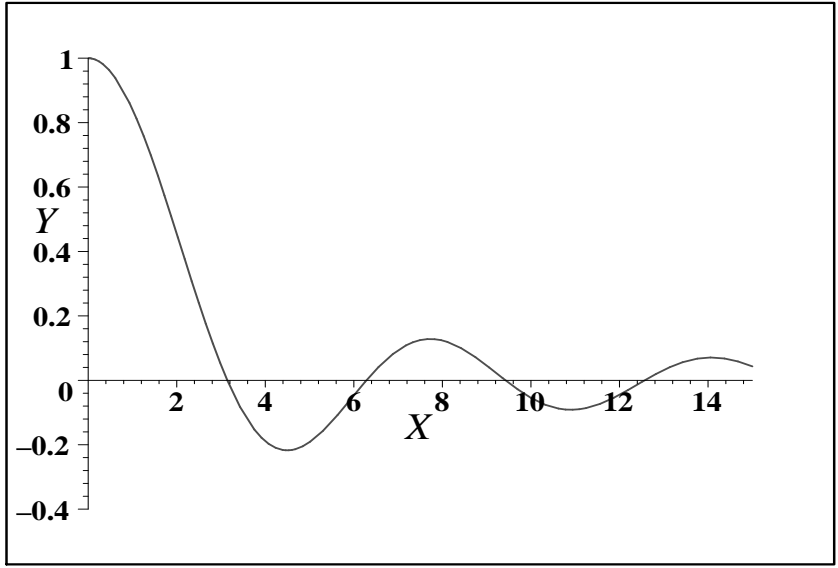,height=7.5cm,width=11.0cm}
  \caption{\label{chigr} the coefficient
  $Y=\chi_{n}/(\gamma B_{n})$
  versus the parameter $X=\gamma r$.}
\end{center}
\end{figure}

\begin{figure}[htb]
\begin{center}
  \epsfig{file=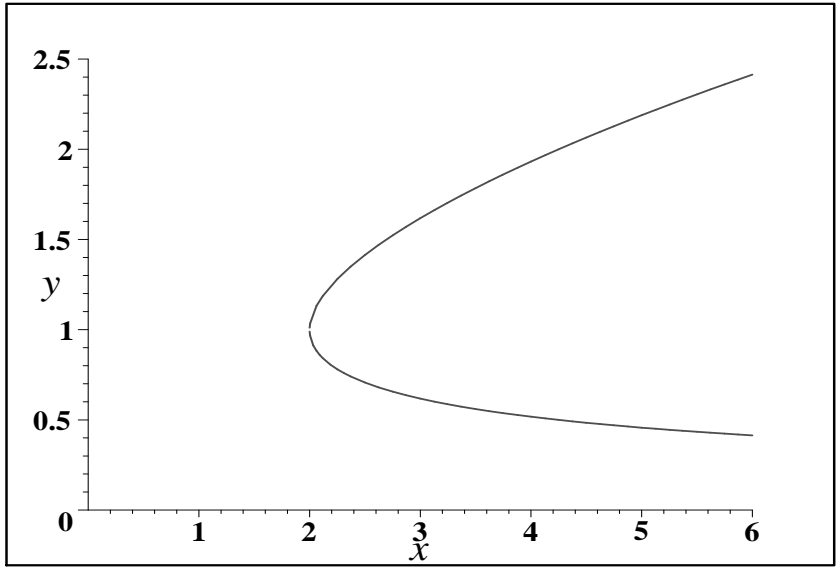,height=7.5cm,width=11.0cm}
  \end{center}
  \caption{\label{disprel} the coefficient
  $y=\gamma[(2\omega_{p}m)/\hbar]^{-1/2}$ versus the parameter
  $x=\omega/\omega_{p}$. If $\omega=2\omega_{p}$, then
  $\gamma_{1,2}= \pm [(2m\omega_{p})/\hbar]^{1/2}$.
  If $\omega \gg 2\omega_{p}$, for the upper branch we have
  $\gamma_{1,2}\sim\pm[(2m\omega)/\hbar]^{1/2}$,
  and for the lower branch
  $\gamma_{3,4}\sim\pm[(2m\omega_{p}^{2})/(\hbar\omega)]^{1/2}$.}
\end{figure}

For the total linearization, it is necessary to suppose that
$2\Psi _{0} + f \approx 2\Psi _{0}$ in the third term of the
Eq.~(\ref{eqf}). Then, we represent the function $\chi$ through
its real and imaginary parts: $\chi=\chi_{1}+i\chi_{2}$. Hence,
$f=\chi_{1}\cos\omega t+\chi_{2}\sin\omega t$, and the
Eq.~(\ref{eqf}) can be divided into two independent similar
equations for $\chi_{1}$ and $\chi_{2}$:
\begin{equation}
  \label{eqchi}
  \hbar \omega  \chi _{n} +
  \frac{{\hbar ^{2}}}{{2m}}\Delta  \chi_{n} -
  8\pi e^{2}n_{0}F\left( {\chi_{n}}  \right)=0,
  \quad n = 1,2.
\end{equation}
One can find out that the functions
$$
  \chi _{n} = B_{n}\frac{{\sin\gamma r}}{{r}}, \quad
  n=1,2,
$$
where $B_{n}$ are real constants, are the solutions of
the Eq.~(\ref{eqchi}). The behavior of the functions
$\chi_{n}$ is shown in Fig.~\ref{chigr}.
The parameter $\gamma$ satisfies the dispersion relation
\begin{equation}
  \label{gamrel}
  \gamma^{2}={\frac{\omega m}{\hbar}}
  \left[
  1\pm
  \left(
  1-4{\frac{\omega_{p}^{2}}{\omega^{2}}}
  \right)^{1/2}
  \right],
\end{equation}
where $\omega_{p}$ is determined in the Eq.~(\ref{plfr}). The
positive part of the Eq.~(\ref{gamrel}) as a function of frequency
$\omega$ is presented in Fig.~\ref{disprel}.
It is worth
noticing that if the values of $B_{n}$ are limited, the
expressions for $\chi_{n}$ satisfy the condition of system
electroneutrality Eq.~(\ref{eln}).

Now we present the estimate of some characteristics of the
oscillations process. For the density $n_{0}=2.7\times 10^{19}{\rm
cm}^{-3}$, i.e. for completely singly ionized gas under the
atmospheric pressure and when $\omega=2\omega_{p}$, the frequency
of electron oscillations is:
$$
  \nu={\frac{\omega_{p}}{\pi}}= 2\left( {\frac{e^{2}n_{0}}{\pi m}}
  \right)^{1/2} \approx 9\times 10^{13}{\rm Hz}.
$$
This frequency corresponds to the electromagnetic radiation in the
infrared range with the wavelength $\lambda={{c}/{\nu}}\approx
3\times 10^{-4}{\rm cm}$. In this case
$\gamma_{1,2}=\pm[(2m\omega_{p})/\hbar]^{1/2} \approx\pm 2.3\times
10^{7}{\rm cm}^{-1}$. The size of the central region $\delta$,
where the most intensive oscillations of the electron gas are
observed, is equal to $\pi / \gamma \approx 1.4\times 10^{-7} {\rm
cm}$. The obtained value for $\delta$ is of atomic order. That is
why we have adopted the quantum approach to the electron gas
oscillations problem.

From the Eq.~(\ref{gamrel}) one can see that the frequency
$\omega=2\omega_{p}$ is the critical value, since for frequencies
less than $2\omega_{p}$, $\gamma$ becomes a complex value and
under these circumstances the oscillations of the electron gas are
damped. However, there is no contradiction between this result and
the well known fact that in using the classical approach to
similar problem one gets the value $\omega=\omega_{p}$ for the
critical frequency. The plasma frequency $\omega_{p}$ is the
critical value for a frequency in a sense that oscillations of
electron gas can appear only if $\omega\geq\omega_{p}$. According
to our result $\omega$ should be greater (or equal) than
$2\omega_{p}$. However, under such condition the inequality
$\omega\geq\omega_{p}$ is sure to be satisfied. The obtained
constraint for possible oscillations frequencies $\omega\geq
2\omega_{p}$ is unlikely to be the feature of geometry of the
system in question. As it will be shown below, there is the same
condition for axially-symmetrical electron gas oscillations. We
suspect that the deviation from the result of classical theory
($\omega\geq\omega_{p}$) results from the quantum approach used in
our paper.

Now let us discuss the origin of each of the branches in the
dispersion relation. At high oscillations frequencies the
parameter $\gamma$ is great for the upper branch. Therefore, we
can define the considered branch as the "high-energy" one.
Moreover, it is possible to treat the upper branch as the
"classical" one. If $\omega\gg 2\omega_{p}$, the dispersion
relation for this branch becomes similar to the relation between
energy $\hbar\omega$ and "momentum" $\hbar\gamma$ for a classical
particle:
$$
\hbar\omega=\frac{(\hbar\gamma)^{2}}{2m}.
$$
Note that the definition "high-energy" is consistent with
"classical" owing to the correspondence principle between
classical and quantum mechanics (see, for instance,
Ref.~\cite{La.Li/QM(89)}).

The lower branch is the antithesis to the upper branch in all
respects. To begin with the behavior of the parameter $\gamma$ at
high frequencies. Contrary to the upper branch the lower one
vanishes as $\omega^{-1/2}$. This fact means that we can define
considered branch as the "low-energy" one. Thus, according to the
correspondence principle again we can also treat it as the
"quantum" branch. We suspect that such substantially quantum
phenomenon as superconductivity of electron gas may take place if
the dispersion relation is realized as the lower branch. This
problem is also discussed in Sec.~\ref{APP} devoted to possible
applications of the described model.

Now let us discuss the ion gas density distribution. We supposed
that number density of the ion gas was constant throughout the
volume because of the small mobility of heavy ions. However, it
leads to incorrect results. It is necessary to remind that the
exact expression for the density of the electron gas in searching
for the solution in the form of the Eq.~(\ref{form}) is presented
as $\left| \Psi \right|^{2}= n_{0} + \Psi_{0} f +\left| \chi
\right|^{2}$. Let us define in this expression $\Psi_{0} f$ and
${\bar n}_{e}=n_{0}+\left| \chi \right|^{2}$ as the dynamic and
static components of the electron gas density. In deriving the
approximate linearized Eq.~(\ref{eqchi}) the function $\left| \chi
\right|^{2}$ was supposed to be small and thus neglected. This
procedure is not correct because the integral operator $F\left(
{\left| \chi \right|^{2}} \right)$ becomes divergent for the
function $\chi=(B/r)\sin\gamma r$. The divergence is caused by the
time independent quantity $|\chi|^{2}$, which means the excessive
static component of the electron gas density. We assumed that the
density of the ion gas was constant. The average of the frequently
oscillating dynamic component over time is zero, but $\left| \chi
\right|^{2}$ does not depend on time. It is naturally to expect
that under some conditions the negative volume charge, described
by the function $|\chi|^{2}$, will be compensated (or neutralized)
by the removing of positive ions. This process will result in
local changing of density and pressure of the ion gas.

Indeed, it can be shown that for our case the condition of static
stability of the ion gas is:
\begin{equation}
  \label{intstab}
  kT
  {\frac{\partial n_{i}}{\partial r}}=
  {\frac{4\pi e^{2}}{r^{2}}}
  n_{i}
  \int\limits_{0}^{r}
  \left( n_{i} - {\bar n}_{e} \right) x^{2}dx
\end{equation}
where $k$ is Boltzmann constant, $T$ is the ion gas temperature,
$n_{i}$ is the ions density.

For the neutralization process to occur we demand that the ions
density $n_{i}$ should be equal to the static electrons density
${\bar n}_{e}=n_{0}+|\chi|^{2}$ with high level of accuracy. Then,
from Eq.~(\ref{intstab}) we get the following inequality:
$$
  \frac{kTr^{2}}{4\pi e^{2}n_{i}}
  \left|\frac{\partial n_{i}}{\partial r}\right|\ll 1.
$$
Having substituted the expression for $n_{i}$ in the last formula,
we obtained the condition of the neutralization:
$(kTB^{2}\gamma)/(4\pi e^{2}n_{0})\ll 1$. Taking into account that
ions density in the center of the system should be equal to
$n_{c}=B^{2}\gamma^{2}$, this condition can be rewritten in the
following way:
\begin{equation}
  \label{neutr}
  n_{c}\ll {\frac{4\pi e^{2}\gamma}{kT}}n_{0}.
\end{equation}
For instance, for $T=10^{3} {\rm K}$ and the value
$\gamma=2.3\times 10^{7} {\rm cm}^{-1}$, which was obtained above,
we get that $n_{c}\ll 500 n_{0}$. We supposed that the
perturbations described by the function $\chi$ were small, i.e.
$n_{c}\ll n_{0}$. Hence, the condition Eq.~(\ref{neutr}) is
satisfied and the excessive static electron charge is undoubtedly
compensated by the ion charge and the divergence in the integral
$F(|\Psi|^{2}-n_{i})$ can be eliminated. Thus, the approximate,
linearized theory trends to describe the pressure enhancement of
the ion gas in the center of symmetrically oscillating electron
gas.

The ion density enhancement, described in our paper, and the Debye
shielding are the completely different phenomena since the latter
is the static effect and the former is the dynamic phenomenon. It
should be also noted that the electron gas density enhancement is
the essentially quantum effect and cannot be obtained within the
classical approach with the use, for example, of the kinetic
equation method. Indeed, in the classical description of the
electron gas oscillations in plasma we get that electron
distribution function in the Eq.~(\ref{kineq}) has the form:
$$
  f_{e}({\bf r},{\bf p},t)=
  f_{0}({\bf p})+\delta f({\bf r},{\bf p},t),
$$
where $f_{0}({\bf p})$ is the unperturbed stationary as well as
spatial coordinates independent distribution function, $\delta
f({\bf r},{\bf p},t)$ is the perturbation of the distribution
function. The electrons number density is expressed in the
following way
$$
  n_{e}({\bf r},t)=\int d^{3}{\bf p}f_{e}({\bf r},{\bf p},t)=
  n_{0}+\delta n({\bf r},t),
$$
where
\begin{equation}
  \label{deltan}
  \delta n({\bf r},t)=
  \int d^{3}{\bf p}\delta f({\bf r},{\bf p},t).
\end{equation}
The function $\delta f({\bf r},{\bf p},t)$ is usually taken to be
proportional to $\exp[i({\bf k}{\bf r}-\omega t)]$. Therefore, the
average of $\delta n({\bf r},t)$ over time in the
Eq.~(\ref{deltan}) is zero. Thus, it is impossible to obtain the
electron gas density enhancement within the context of the
classical theory.

Basing on the obtained approximate solutions of the non-linear
Shr{\"o}dinger equation, we suggest the following dynamics of the
spherically symmetrical electron gas oscillations with {\it
finite} amplitude:
\begin{enumerate}
\item
The appearance of the excessive negative volume charge in the
center of the system (quantum mechanical effect in electron gas
oscillations).
\item
The neutralization of the small negative volume
charge as a result of the ions motion towards the center.
\item
The
enhancement of the ion gas density and pressure.
\item
The Hamiltonian
in the Eq.~(\ref{ham}) is changed allowing for the new electric
charges distributions. The approximate solution of the new equation
provides further pressure enhancement of the ion gas.
\end{enumerate}

While considering non-linear Shr{\"o}dinger equation
(\ref{trdifeq}), it can be seen that along with the components on
frequency $\omega$, the terms which do not depend on time as well
as on frequencies $2\omega$, $3\omega$ and etc appear. One can
make sure of this representing the solution of the
Eq.~(\ref{trdifeq}) in the form:
$$
  \Psi (r,t)=\Psi_{0}+\sum\limits_{k=1}^{\infty} \Psi_{k}(r,t),
$$
where $\Psi_{1}=(B_{1}/r)e^{-i\omega t}\sin \gamma r$.

It is worth mentioning that along with spherically symmetrical
solution of the Eqs.~(\ref{difeq})-(\ref{thet}),
there is at least one
axially-symmetrical solution which has the form:
$$\Psi (r,t) = n_{0}^{1/2}+B
J_{0}(\gamma r) e^{-i\omega t},$$
where $J_{0}$ is the zero-order Bessel function, $B$ is a real
constant. In this case the dispersion relation takes the same form
as the Eq.~(\ref{gamrel}). All consequences obtained for
spherically-symmetrical oscillations are valid for this case as
well.

\section{\label{APP} Applications}

In this section we present the data of the {\it fireball}
appearance in nature and various theoretical models for the
description of this phenomenon. We also analyze the possibility
for a {\it fireball} to be accounted for in frames of our model of
quantum electron gas oscillations in plasma.

The {\it fireball} is a very rare natural phenomenon. However,
numerous observations of fireballs have been collected (see, for
example, Ref~\cite{St/PNF(85)}). It should be noted that most of
the fireball observers were not professional researchers and,
thus, their descriptions of this phenomenon may appear to be quite
subjective. However, the substantial number of these observations
reveal some regularity. We present below the key aspects of the
fireball observations.
\begin{itemize}
  \item {\bf The fireball appearance.} A {\it fireball} more often
  appears beside the lightning stroke. There are, however, a
  considerable amount of observations when a {\it fireball}
  appeared without a lightning. Different sharpened objects such as
  metallic pieces, antennae, wires etc are reported to favor the
  {\it fireball} appearance. The {\it fireball} observers usually
  mention that there can be the high electric field strength, for
  instance, corona discharge on the metallic objects.
  \item {\bf The fireball disappearance.} A {\it fireball} often
  disappears smoothly attenuating. However, rather frequently a
  {\it fireball} disappears with explosion, releasing of great
  amount of heat energy and melting metal and sand.
  \item {\bf Stationary state of a fireball} The usual average
  size of a {\it fireball} is about $30\div 40 {\rm cm}$. The
  lifetime is in the range from $1$ to $200 {\rm s}$. The color of
  the {\it fireball} glow is various: red, yellow, white and blue.
  The minimal number of observed {\it fireballs} had green hues.
  Neither high brightness of a {\it fireball} nor the effects
  which can be regarded as the evidence of thermal radiation are
  the characteristics of a {\it fireball}. Its light looks like a
  glow discharge. Even transparent {\it fireballs}
  are reported to appear.
  Sometimes the observers mention that a {\it fireball} has more
  bright core.
  \item {\bf The fireball form.} A {\it fireball} often has
  elliptical or spherical form. The form, which can be identified
  with toroidal one, is rarely observed. The {\it fireball}
  surface is reported to consist of needles. The {\it fireball}
  sparking is the frequently mentioned feature.
  \item {\bf The fireball motion and location.} A {\it fireball} is
  regularly located near objects on the ground or in the
  atmosphere (for example, near airplanes). It moves, as a rule,
  along these objects on a certain distance. A {\it fireball} can
  be elastically reflected by the objects or disappears in touch
  with them. A {\it fireball} moves towards the air current as
  well as in the reversed direction. A {\it fireball} can
  penetrate indoors through a chimney, window glass and holes,
  with their diameter being less than the visible one of a
  {\it fireball}. It is worth noticing that window glass is not
  destroyed sometimes when a {\it fireball} travels through it.
  The {\it fireball} interaction with ferromagnetic materials is
  not observed.
  \item {\bf The biological action of a fireball.} The biological
  effects of a {\it fireball} are the legs and arms paralysis as
  well as the tan.
\end{itemize}

The {\it fireball} models can be divided into several groups.
According to one of them a {\it fireball} appears if the
electromagnetic radiation with the wavelength of several
centimeters or shorter is closed on itself forming a stable wave
package. This process occurs either in the atmosphere or on the
ground in a certain form a lightning discharge. Other {\it
fireball} models are based on the step leader. A {\it fireball} is
either identified with the step leader or is its extension under
certain conditions ("unfinished" lightning, combination of
currents values etc).

According to the model of quantum electron gas oscillations in
plasma, developed in this paper, we found that if one presents the
solution of the non-linear Shr{\"o}dinger equation in the form of
the Eq.~(\ref{form}), then time independent excessive component of
the electron gas density always appears. It leads to the ion gas
density enhancement in the center of the system. This property is
unlikely to be the feature of the adopted Hamiltonian.

In order to maintain plasma in ionized state and realize the
process described above, the energy ingress from the outside or
the energy release within the system is wanted. Nuclear fusion
reactions can, in principle, serve as similar source of energy.
The reactions will proceed if pressure and density of the ion gas
attain appropriate values.

This process seems to support the existence of a {\it fireball}.
Water is known to contain deuterium in the amount of $\approx
1/5000$ under normal conditions. If water vapors are present in
atmosphere, the running the nuclear fusion reactions will release
energy, which supports the oscillations of the electron gas and
prevents the recombination of plasma. Possible exothermic nuclear
fusion reactions are:
$$
{^{2}}{\rm D}_{1}+{^{2}}{\rm D}_{1}\to {^{3}}{\rm He}_{2}+
{^{1}}{\rm n}_{0},
$$
$$
{^{2}}{\rm D}_{1}+{^{2}}{\rm D}_{1}\to {^{3}}{\rm T}_{1}+
{^{1}}{\rm p}_{1},
$$
$$
{^{2}}{\rm D}_{1}+{^{3}}{\rm T}_{1}\to {^{4}}{\rm He}_{2}+
{^{1}}{\rm n}_{0},
$$
where ${^{2}}{\rm D}_{1}$ and ${^{3}}{\rm T}_{1}$ are the heavy
isotopes of hydrogen (deuterium and tritium), ${^{3}}{\rm He}_{2}$
and ${^{4}}{\rm He}_{2}$ are the helium isotopes, ${^{1}}{\rm
n}_{0}$ and ${^{1}}{\rm p}_{1}$ are the neutron and the proton.
Taking into account small sizes of the central (active) region and
small amount of deuterium in atmosphere, it is possible to use the
term 'microdose' nuclear fusion reactions for the process in
question. Axially-symmetrical oscillations of the electron gas are
likely to appear as very seldom observed type of a {\it fireball}
in the form of shining, sometimes closed cord (see also the list
of the observed {\it fireball} properties in the beginning of this
section). Uncomplicated calculation shows that energy released in
deuterium nuclei fusion in $1{\rm dm}^{3}$ of water vapors (the
average size of a {\it fireball}) has the value of about $1{\rm
MJ}$, that corresponds to energy evaluations of some observed {\it
fireballs} \cite{St/PNF(85),Fr/ZETF(40)}.

However, along with high-energy 'fireballs' there often appear
low-energy ones. The energy estimate of such 'fireballs' (see
Ref.~\cite{St/PNF(85)}) indicates that the nuclear fusion
reactions are unlikely to support their existence. Low-energy
{\it fireball} may be presented in frames of our model as the solution
with the dispersion relation described by the lower branch (see
Fig.~\ref{disprel}). The superconductivity might be the mechanism
preventing possible attenuation of the electron gas oscillations
caused by various dissipation processes.

Moreover, the facts, indirectly verifying that quantum electron
gas oscillations in plasma are the possible model of a {\it
fireball}, are (see also the list of the observed {\it fireball}
properties in the beginning of this section):
\begin{enumerate}
  \item Numerous {\it fireball} observers mention the elevated
  level of the atmospheric gas ionization when a {\it fireball}
  appears. This fact points out that a {\it fireball} generates
  very strong electromagnetic radiation with the frequency in the
  ultraviolet range (or even higher). Although the systems with
  spherically-symmetrical electron gas oscillations do not reveal
  radiation, the ultraviolet electromagnetic radiation could be
  generated in the outer layers of such systems. The origin of
  this radiation would be various secondary effects, for example,
  small deviations from spherically-symmetrical state.
  \item Persons who were beside a {\it fireball} sometimes
  revealed the tan. The explanation of this fact is the same as in
  the item 1.
  \item A {\it fireball} often disappears with heavy explosion.
  This point indicates that there is the elevated pressure region
  within a {\it fireball}. This region should have small size. If
  the size had been great, the weight of a {\it fireball} would
  have exceeded the weight of air and a {\it fireball} would have
  descended to the ground. However, it confronts the observations
  that a {\it fireball} freely moves in the atmosphere. The
  pressure and density of the ion gas are inevitably increased
  if the electron gas oscillations in plasma are described in
  frames of our model.
  \item A {\it fireball} burns small holes, from several
  millimeters to $2\div 3 {\rm cm}$ across, when it travels
  through dielectric materials (glass, plastic) as well as through
  thin metallic sheets. If a {\it fireball} is described
  on the basis of our model,
  the nuclear fusion reactions may take place in its central
  region. Thus, high temperature, required to melt, for example,
  glass, can be achieved.
  \item The {\it fireball} lifetime is anomalously high. This fact
  can be accounted for by none of the models based on the
  classical electrodynamics or hydrodynamics.
\end{enumerate}

\subsection{Experimental studying of electron gas oscillations in plasma}

Groups and separate researchers developing the problem of the
controlled fusion are suggested to pay attention to
self-consistent, radially and axially oscillating electron
'plasmoids' as a base models to self-supported nuclear fusion
reactions. The authors of this article have certain experience in
generating ball-like plasma structures. We present below the
description of the experiments carried out in the N.~N.~Andreev
Institute of Acoustics in 1995-1999.

\begin{figure}[htb]
\begin{center}
\epsfig{file=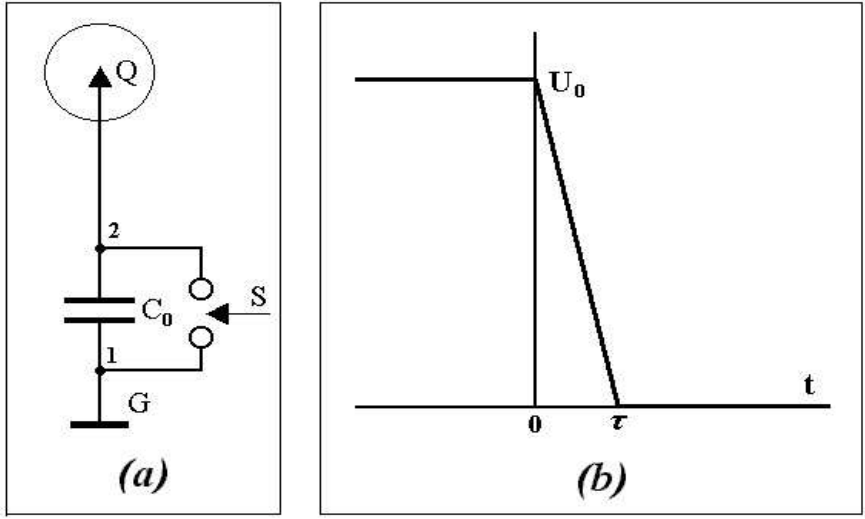,height=7cm,width=11.6cm}
\end{center}
\caption{\label{pl1} (a) the circuit for the generation of
spherical plasma structures, (b) the process of the capacitor
$C_{0}$ discharge.}
\end{figure}

The plant, used to generate spherical plasma structures, is shown
in Fig.~\ref{pl1}(a).
The reservoir capacitor $C_{0}\approx 10^{-7}{\rm F}$ was charged
up to the voltage $U_{0}\approx 2\times 10^{5}{\rm V}$. The
electrostatic field energy of the capacitor was
$W_{el}=\frac{C_{0}U_{0}^{2}}{2}\approx 2\times 10^{3}{\rm J}$.
After the complete capacitor discharge, the point $Q$ formed very
strong corona charging about $5{\rm cm}$ across. The corona was
visible even in diffuse sunlight. The point $Q$ was the sharpened
quenched steel wire. The space near the point was enriched with
heavy water (${\rm D}_{2}{\rm O}$). When a high-voltage pulse was
applied to the electrode $S$, the break-down of the spark gap
occurred and the capacitor $C_{0}$ was short-circuited. The
process of the capacitor discharge is schematically presented in
Fig.~\ref{pl1}(b). The voltage fell from $200{\rm kV}$ to zero
during the time $\tau\approx 10^{-8}{\rm s}$. Glowing dark-red
spheres about $3{\rm cm}$ across separated from the point $Q$ at
the moment of the break-down. The lifetime of these structures was
in the interval $1\div 5{\rm s}$.

The described discharge scheme seems to have implemented the
excitation of the spherically symmetrical electron gas
oscillations. The electron gas was formed by the high-voltage
corona discharge on the point $Q$.

\begin{figure}[htb]
\begin{center}
\epsfig{file=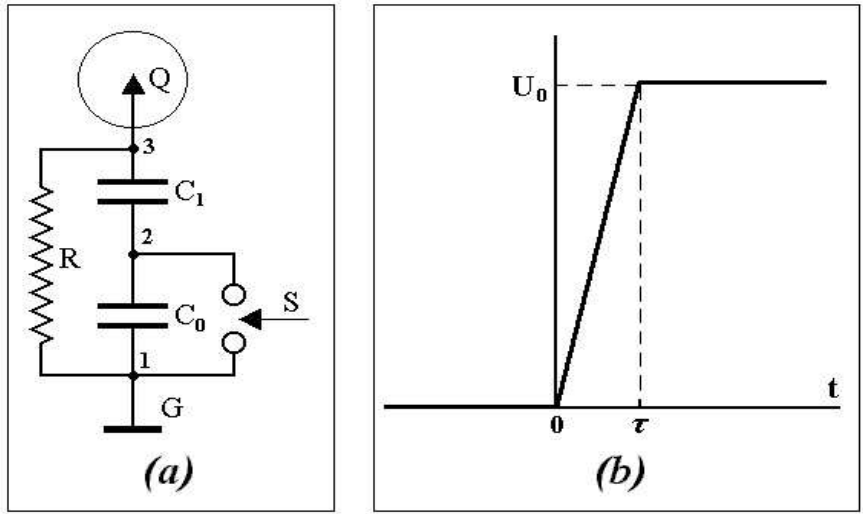,height=7cm,width=11.6cm}
\end{center}
\caption{\label{pl2} (a) the modified circuit for the generation of
spherical plasma structures, (b) the potential of the point $Q$.}
\end{figure}

The disadvantages of the considered plant are:
\begin{itemize}
\item
high power of the charging devise caused by strong leakage current
in displaying a corona discharge;
\item
high level of noise in short-circuiting of the capacitor $C_{0}$;
\item
big overall dimensions
of the plant.
\end{itemize}
The listed disadvantages were partially removed in the plant which
is shown in Fig.~\ref{pl2}(a).
In this circuit, $C_{1}$ denotes the reservoir capacitor. When
high voltage $U_{0}$ was applied to the point $2$ of the circuit,
the potential of the point $Q$ was always equal to zero because of
the shunting resistor $R$. After the break-down of the spark gap,
the potential of the capacitor $C_{0}$  fell to zero and the
potential of the point $Q$ became equal to $U_{0}$. This process is
schematically depicted in Fig.~\ref{pl2}(b). The parameters of
this circuit were: $C_{0}\approx 10^{-10}{\rm F}$, $C_{1}\approx
10^{-9}{\rm F}$, $R\approx 10^{9} \Omega$, $U_{0}\approx 1.5\times
10^{5}{\rm V}$. Leakage currents were almost eliminated since all
reservoir elements had been placed in transformer oil. The principal
elements of this circuit have passed preliminary testing.

At the end of this section we briefly discuss the results of the
experiments, carried out for the purpose to study the interaction
of laser radiation with the tantalum targets, which are presented
in Ref.~\cite{Sk.Vo/EWES(02)}. The authors of this paper report
that long-lived structures were produced in the experiments. These
objects are reported to emit hard gamma-ray radiations (with
energy of quanta $E\geq 0.1 {\rm MeV}$) and abandon specific
tracks on the metallic walls of the vacuum chamber. These facts
imply that, as it was also mentioned by the authors of
Ref.~\cite{Sk.Vo/EWES(02)}, the density of matter within the
generated structures is very high. Such stable, long-lived
structures cannot be described be means of classical
electrodynamics. This point was also noticed in
Ref.~\cite{Sk.Vo/EWES(02)}. We suppose that the authors of this
article have exited the electron gas oscillations which may be
described in frames of our model. Thus, for example, the elevated
pressure in these objects can be successfully accounted for. The
suggestion that the superconductive electric current might
circulate within the produced structures was put forward in
Ref.~\cite{Sk.Vo/EWES(02)}. This hypothesis agrees with our
anticipation that superconductivity might occur in the outer
layers of spherically and axially symmetrical electron gas
oscillations in plasma.

\section{\label{CONCL} Conclusion}

The quantum mechanical description of the electron gas
oscillations in plasma has been presented in this paper.  We have
analyzed spherically and axially symmetrical solutions of the
non-linear Shr{\"o}dinger equation. The conditions of the
oscillations process existence have been examined. It has been
found that in the center or on the axis of symmetry of the systems
the static density of electrons enhanced. This process led to the
increasing of density and pressure of the ion gas. We have also
discussed possible applications of the obtained solutions. It has
been suggested that this mechanism could occurs in nature as rare
phenomenon called the {\it fireball} and could be used in carrying out
the research concerning controlled fusion.  The description of the
experiments, carried out for the purpose to generate long-lived
spherical plasma structures, has been presented.

New approaches to the technical development of the controlled
fusion problem are requested nowadays. The collected experience of
plasma physics allows one to operate with very complicated plasma
configurations. The problem of the self-consistent plasma
confinement in power-generating plants can probably be imposed. In
this situation a {\it fireball} is of particular interest because of
its high energy and long lifetime. A great number of the
observations concerning a {\it fireball} appearance was collected.
Plenty of the {\it fireball} models has been constructed. The advances
in the studying of plasma interactions with rf radiation and
matter point out that the stable plasma structures, supplied by
the energy of nuclear fusion reactions, can, in principle, be
implemented.

Taking into account limited amount of the exhaustible resources of
natural energy carriers, the mankind will inevitably face energy
catastrophe in 21st century. The consequence will be the complete
destruction of the modern industrial civilization. The alternative
energy sources, such as geothermal, tidal, wind-power, solar and
water-power stations, will be unable to prevent this catastrophe.
The only possibility for the mankind is to acquire the energy
released in the controlled fusion, using for this purpose almost
inexhaustible deuterium resource in the oceanic water. However,
the numerous attempts to implement the stable controlled fusion,
using the technique of electrical heating of plasma with its
subsequent squeezing by the external magnetic field, are likely to
fail. Now it is possible to claim that this technique is a
dead-end for the outlined problem. In our opinion, the
self-consistent plasma structure - a {\it fireball}, is one of the
possible ways to acquire almost inexhaustible energy resource.

\end{document}